\documentstyle[11pt,epsf]{article}
\newtheorem{theorem}{Theorem}

\newcommand {\dfn} {\stackrel{\Delta} {=}}
\newcommand {\exe} {\stackrel{\cdot} {=}}
\newcommand {\lexe} {\stackrel{\cdot} {\le}}

\newcommand {\reals} {{\rm I\!R}}

\newcommand {\bx} {\mbox{\boldmath $x$}}

\newcommand {\by} {\mbox{\boldmath $y$}}

\newcommand {\bE} {\mbox{\boldmath $E$}}

\newcommand {\bX} {\mbox{\boldmath $X$}}

\newcommand{\calA}{{\cal A}}

\newcommand{\calC}{{\cal C}}

\newcommand{\calE}{{\cal E}}

\newcommand{\calX}{{\cal X}}
\newcommand{\calY}{{\cal Y}}

\begin{document}
\thispagestyle{empty}
\title{Another Look at Expurgated Bounds and Their\\ Statistical--Mechanical
Interpretation\thanks{
This research was supported by the Israeli Science Foundation (ISF) grant no.\
412/12.}}
\author{Neri Merhav
}
\date{}
\maketitle

\begin{center}
Department of Electrical Engineering \\
Technion - Israel Institute of Technology \\
Technion City, Haifa 32000, ISRAEL \\
E--mail: {\tt merhav@ee.technion.ac.il}\\
\end{center}
\vspace{1.5\baselineskip}
\setlength{\baselineskip}{1.5\baselineskip}

\begin{abstract}
We revisit the derivation of expurgated error exponents using a method of type
class enumeration, which is inspired by statistical--mechanical methods, and which has
already been used in the derivation of random coding exponents in several
other scenarios. We compare our version of the expurgated bound to both the
one by Gallager and 
the one by Csisz\'ar, K\"orner and Marton (CKM). For expurgated ensembles of fixed
composition codes over finite alphabets, our basic expurgated bound coincides with the CKM
expurgated bound, which is in general tighter than Gallager's bound, but with
equality for the optimum type class of codewords. Our method, however,
extends beyond fixed composition codes and beyond finite alphabets, 
where it is natural to impose input
constraints (e.g., power limitation).
In such cases, the CKM expurgated bound
may not apply directly, and our bound is in general
tighter than Gallager's bound. In addition, while both the CKM and the Gallager
expurgated bounds are based on Bhattacharyya bound for
bounding the pairwise error
probabilities, our bound allows the more
general Chernoff distance measure, thus giving rise to additional improvement using 
the Chernoff parameter as a degree of freedom to be
optimized.\\

\end{abstract}

\newpage
\section{Introduction}

It is well known that the random coding exponent on the probability of error in
channel coding can be improved, at low coding rates, by a process called {\it
expurgation}, that results in the so called {\it expurgated exponent}, or
the {\it expurgated bound}, which is a lower bound to the reliability function. The
idea of expurgation, first introduced by Gallager \cite[Section
V]{Gallager65}, \cite[Section 5.7]{Gallager68} (see also \cite[Section
3.3]{VO79}), is that at low rates, the average error probability over the
ensemble of codes, is dominated by bad randomly chosen codewords and not by
the channel noise, therefore, by eliminating some of these codewords (while keeping the
rate almost the same), an improved lower bound 
on the reliability function is obtained. The expurgated bound at
zero rate is known to be tight, as it coincides, at this point, with the
straight--line bound, which is an an upper bound on the reliability function
\cite[Section 5.8]{Gallager68}, \cite{SGB67a}, \cite{SGB67b}, \cite[Sections 3.7, 3.8]{VO79}.
Omura \cite{Omura74} was the first to relate the expurgated exponent at low
rates to distortion--rate functions, where the Bhattacharyya distance function plays
the role of a distortion measure. 

Several years later,
Csisz\'ar, K\"orner and Marton \cite{CKM77} derived, for finite alphabets, a different expurgated
bound, henceforth referred to as the
{\it CKM expurgated exponent}, as opposed to the
{\it Gallager expurgated exponent} discussed above. 
While ref.\ \cite{CKM77} contains no details (it is an abstract only), the CKM
expurgated exponent is mentioned in \cite[eq.\ (7)]{Csiszar82} and some hints on its
derivation can be found in \cite[p.\ 185, Problem 17]{CK81}. While the CKM
expurgated exponent is equivalent to that of Gallager for the optimum channel input
assignment \cite[p.\ 193, Problem 23(b)]{CK81}, it turns out (as we will be
shown below) that for a
general input distribution,
the CKM expurgated bound is larger (and hence tighter) than the Gallager expurgated bound.
This is important whenever channel input constraints (e.g., power limitation)
do not allow this optimum input distribution to be used.
On the other hand, since the derivation \cite[pp.\ 185--186, Problem 17 (hint)]{CK81} 
of the CKM expurgated exponent 
relies strongly on the packing lemma
\cite[p.\ 162, Lemma 5.1]{CK81}, it is limited to finite input and output
alphabets (as mentioned) and to fixed composition codes, as opposed to the
Gallager expurgated exponent, whose derivation is carried out under more general conditions.

In this paper, our quest is to enjoy the best of both worlds: We use yet
another analysis technique, which has already been used in several previous
works in different scenarios \cite{p121}, \cite{p122}, \cite{p117},
\cite{p120}, \cite[Chapters 6,7]{p138}, \cite{SM12}, \cite{p123}, where it has
always yielded simplified and/or improved bounds on error exponents. This
technique, which is based on distance enumeration, or more generally, on type class enumeration,
is inspired by the statistical--mechanical perspective on random coding, based
on its analogy to the random energy model \cite[Chapters 5, 6]{MM09}, 
which is a model of spin glasses with a high degree of disorder, 
invented by Derrida
\cite{Derrida80a}, \cite{Derrida80b}, \cite{Derrida81},
and which is well known in the literature
of statistical physics of magnetic materials. Our technique is applicable to
channels with quite general input/output alphabets, it is not limited to fixed composition
codes, and it allows the incorporation of channel input constraints, which
are, of course, especially
relevant when the channel input alphabet is
continuous. In the
special case of finite alphabets, our basic 
bound coincides with the CKM expurgated bound
along the whole interesting range of rates, and hence is tighter, in general, than Gallager's
expurgated exponent. 

Furthermore, an additional improvement of our expurgated bound is obtained by
observing that, instead of using the Bhattacharyya bound for the pairwise 
error probabilities (as is done in the derivations of both the Gallager- and the CKM expurgated
exponents), it turns out that
for our proposed form of the expurgated exponent, the pairwise error
probabilities can more generally be bounded using the Chernoff 
distance measure, whose parameter is subjected to optimization.\footnote{While
Gallager's bound has a symmetry that guarantees that the optimum value of the Chernoff
parameter is always $1/2$ (in which case, the Chernoff distance coincides with
the Bhattacharyya distance), this symmetry does not appear in the new proposed bound,
and hence the optimum value of the Chernoff parameter is not necessarily
$1/2$.}

Finally, as mentioned above, our analysis technique is based on a statistical--mechanical
point of view. This point of view naturally suggests a physical interpretation
to the behavior of the expurgated exponent in the following sense:
Similarly as in Gallager's and the CKM expurgated exponents, the graph of the
new proposed expurgated exponent is curvy at low rates and becomes a
straight line of slope $-1$ at the higher range of rates. It turns out that
this passage from a curve to a straight line can be understood as a {\it phase
transition} in the analogous statistical--mechanical system model -- the
random energy model. This point will be discussed as well.

The outline of the remaining part of this paper is as follows. In Section 2,
we provide some background on the expurgated exponents of Gallager and
Csisz\'ar, K\"orner and Marton, as well as the relationship between them.
In Section 3, we provide a few elementary observations that serve as a basis
for our proposed derivation of the expurgated exponent. In Section 4, we
present the derivation of the new proposed version
of our expurgated error exponent for finite alphabets and fixed composition
codes. In Section 5, we outline the extension of this analysis to continuous
alphabet channels. Finally, in Section 6, we discuss the
statistical--mechanical perspective of our analysis.

\section{Background}

Consider a discrete memoryless channel (DMC), defined by the
single--letter transition probability functions\footnote{Here and throughout
the sequel, ``probability function'' is a common name for a probability mass
function in the discrete alphabet case and a probability density function in the
continuous alphabet case.}
$P=\{p(y|x),~x\in\calX,~y\in\calY\}$,
where $\calX$ and $\calY$ are the input alphabet and the output alphabet,
respectively. Let $Q=\{q(x),~x\in\calX\}$ be a probability function on the
input alphabet $\calX$. 

Gallager's random coding error exponent function is a well known lower bound on the
reliability function of the DMC \cite{Gallager65}, 
\cite[Section 5.6]{Gallager68}, \cite[Section 3.2]{VO79}. It is given by
\begin{equation}
E_r(R)=\sup_{0\le\rho\le 1}\sup_Q [E_0(\rho,Q)-\rho R]
\end{equation}
where
\begin{equation}
E_0(\rho,Q)=-\ln\left(\sum_{y\in\calY}\left[\sum_{x\in\calX}
q(x)p(y|x)^{1/(1+\rho)}\right]^{1+\rho}\right),
\end{equation}
and where here and throughout the sequel, it is understood that for continuous
alphabets, summations are replaced by integrals. This bound is obtained by
analyzing the exponential rate of the average error probability
associated with a randomly chosen code $\calC_n=\{\bx_1,\ldots,\bx_M\}$,
$M=e^{nR}$, $R$ being the coding rate and $\bx_m\in\calX^n$ being the codeword
associated message number $m\in\{1,\ldots,M\}$, where each component of each
codeword is selected independently at random under $Q$.

At low rates, this lower bound on the reliability function can be improved by
expurgating the randomly chosen code. This expurgation is accomplished by
discarding the `bad' half of the codebook, namely, the 
half of codewords whose conditional error probabilities
$$P_{e|m}=\mbox{Pr}\{\mbox{error}|\mbox{message}~m~\mbox{sent}\}$$
are the largest under maximum likelihood (ML) decoding.
Gallager's expurgated exponent function \cite{Gallager65}, \cite[Section 5.7]{Gallager68} 
\cite[Section 3.3]{VO79} is given by
\begin{equation}
\label{EexG}
E_{ex}(R)=\sup_{\rho\ge 1}\sup_Q [E_x(\rho,Q)-\rho R]
\end{equation}
where
\begin{equation}
\label{Ex}
E_x(\rho,Q)=-\rho\ln\left(\sum_{x,x'\in\calX}q(x)q(x')\left[\sum_{y\in\calY}
\sqrt{p(y|x)p(y|x')}\right]^{1/\rho}\right).
\end{equation}
Improvement over $E_r(R)$ is accomplished whenever the coding rate $R$ is small enough such
that the supremum in eq.\ (\ref{EexG}) is achieved (or approached) by values of $\rho$ that are
strictly larger than $1$, as otherwise 
for $\rho=1$, we have $E_x(1,Q)\equiv E_0(1,Q)$.

In \cite{CKM77} (see also \cite[p.\ 185, Problem 17]{CK81} for details), the
following version of the expurgated exponent was presented by 
Csisz\'ar, K\"orner and Marton (CKM) for channels with 
finite input and output alphabets:
\begin{equation}
\label{EexCKM}
E_{ex}(R)=\sup_Q\inf_{\hat{Q}_{XX'}\in\calA(R,Q)}[I(X;X')+\bE d_B(X,X')]-R,
\end{equation}
where $\hat{Q}_{XX'}$ is a generic 
joint probability mass function over $\calX^2$, that
governs both the mutual information and the expectation in the square brackets 
of eq.\ (\ref{EexCKM}),
$$\calA(R,Q)=\{\hat{Q}_{XX'}:~\hat{Q}_X=\hat{Q}_{X'}=Q,~I(X;X')\le R\},$$ 
and $d_B(\cdot,\cdot)$
is the {\it Bhattacharyya distance function}, defined by
\begin{equation}
d_B(x,x')=-\ln\left[\sum_{y\in\calY}\sqrt{p(y|x)p(y|x')}\right].
\end{equation}
In \cite[p.\ 193, Problem 23b]{CK81} it is asserted that the right--hand sides
of eqs.\ (\ref{EexG}) and (\ref{EexCKM}) are equivalent, thus justifying the
common notation $E_{ex}(R)$ for both expressions. Hereafter, to avoid
confusion between the Gallager and the CKM expurgated exponents, 
we will deviate from the customary notation used above, and re--define the
notation $E_G(\rho,Q)$ for $E_x(\rho,Q)$ (where the subscript $G$ stands for
``Gallager''), and accordingly
\begin{equation}
\calE_{G}(R,Q)=\sup_{\rho\ge 1}[E_G(\rho,Q)-\rho R],
\end{equation}
thus, $E_{ex}(R)=\sup_Q\calE_G(R,Q)$. Similarly, we will denote
\begin{equation}
\calE_{CKM}(R,Q)=\inf_{\hat{Q}_{XX'}\in\calA(R,Q)}[I(X;X')+\bE d_B(X,X')]-R,
\end{equation}
thus, $E_{ex}(R)=\sup_Q \calE_{CKM}(R,Q)$. 

While $\sup_Q\calE_G(R,Q)=\sup_Q
\calE_{CKM}(R,Q)$ as mentioned above, it turns out that for a general choice
of $Q$, the functions $\calE_G(R,Q)$ and $\calE_{CKM}(R,Q)$ may differ. In
fact, as we shall see shortly
\begin{equation}
\calE_{CKM}(R,Q)\ge \calE_G(R,Q)
\end{equation}
for an arbitrary input assignment $Q$. This is an important point since the
optimum input assignment $Q^*$, that achieves $E_{ex}(R)$, might be forbidden in
the presence of channel input constraints (e.g., power limitation), and so, in
such a case, the CKM expurgated exponent may be better than the Gallager expurgated
exponent. On the other hand, there are
two advantages to the Gallager expurgated exponent relative to
the CKM expurgated exponent. The first is that,
unlike the case of the CKM bound, its
derivation is not sensitive to the assumption of finite alphabets and 
fixed composition codes.\footnote{In fact, in the case of a continuous input
alphabet, the notion of fixed composition codes does not really exist altogether.}
The second advantage is that the numerical calculation of $\calE_G(R,Q)$ requires
optimization over one parameter only (the parameter $\rho$), whereas the
calculation of $\calE_{CKM}(R,Q)$ seems 
(at least in its present form)
to require optimization over the entire joint
distribution $\hat{Q}_{XX'}$ (which means many parameters for a large input
alphabet) and moreover, this optimization is subjected to complicated
constraints (defined by $\calA(R,Q)$).

\section{Some Preliminary Observations}

Before presenting the proposed alternative derivation of our expurgated
exponent, we pause to offer a few preliminary observations that would hopefully help
to compare $\calE_G(R,Q)$ and $\calE_{CKM}(R,Q)$ and to understand the
relationships between them, as well as their relation to that of the new bound to be
derived. In particular, our first task is to transform the expression of $\calE_{CKM}(R,Q)$
to a form that has the same ingredients as those of $\calE_G(R,Q)$.

We first define the function
\begin{equation}
D_Q(R)=\min_{\hat{Q}_{XX'}\in\calA(R,Q)}\bE\{d_B(X;X')\}.
\end{equation}
Intuitively, the function $D_Q(R)$ is the distortion--rate function of a
``source'' $Q$ (designated by the random variable $X$) 
with respect to (w.r.t.) the Bhattacharyya distortion measure
$d_B(\cdot,\cdot)$, subject to the additional constraint that the ``reproduction
variable'' $X'$ has the same probability distribution $Q$ as the ``source.''
It is easy to see now that
\begin{equation}
\inf_{\hat{Q}_{XX'}\in\calA(R,Q)}[I(X;X')+\bE d_B(X,X')]=\left\{\begin{array}{ll}
D_Q(R)+R & R\le R_1\\
D_Q(R_1)+R_1 & R > R_1\end{array}\right.
\end{equation}
where $R_1$ is $I(X;X')$ for the optimum $\hat{Q}_{XX'}$ 
that minimizes $[I(X;X')+\bE d_B(X,X')]$
across $\calA(\infty,Q)$, or equivalently, $R_1$ is the rate $R$ at which
$D_Q'(R)=-1$, $D_Q'(R)$ being the derivative of $D_Q(R)$ w.r.t.\ $R$.
Thus, we obtain
\begin{equation}
\calE_{CKM}(R,Q)=\left\{\begin{array}{ll}
D_Q(R) & R\le R_1\\
D_Q(R_1)+R_1-R & R_1 < R < D_Q(R_1)+R_1\\
0 & R > D_Q(R_1)+R_1\end{array}\right.
\end{equation}
where we note that the first line is intimately related to \cite[p.\ 194, Problem
24]{CK81}. We observe then that at low rates, $\calE_{CKM}(R,Q)$ has a curvy part
given by $D_Q(R)$, and for high rates it is given by the straight line of
slope $-1$ that is tangential to the curve $D_Q(R)$.

Let us now take a closer look at the distortion--rate function 
$D_Q(R)$, which is the inverse of the
rate--distortion function $R_Q(D)$, defined similarly, and again with the additional
constraint $Q_{X'}=Q$. This rate--distortion function
has the following parametric representation \cite[eq.\ (13)]{p133}:
\begin{equation}
R_Q(D)=-\inf_{s\ge
0}\left[sD+\sum_{x\in\calX}q(x)\ln\left(\sum_{x'\in\calX}q(x')e^{-sd_B(x,x')}\right)\right],
\end{equation}
where the minimizing $s$ is interpreted as the negative local slope of the function
$R_Q(D)$, i.e., $s^*=-R_Q'(D)$, $s^*$ being the minimizer of the r.h.s.
This function can easily be inverted, similarly as in \cite[eqs.\
(15)--(20)]{p129}, to obtain
\begin{eqnarray}
D_Q(R)&=&-\inf_{s\ge 0}\frac{1}{s}
\left[R+\sum_{x\in\calX}q(x)\ln\left(\sum_{x'\in\calX}q(x')
e^{-sd_B(x,x')}\right)\right]\\
&=&\sup_{\rho\ge
0}\left[-\rho\sum_{x\in\calX}q(x)\ln\left(\sum_{x'\in\calX}q(x')e^{-d_B(x,x')/\rho}\right)-\rho
R\right],
\end{eqnarray}
where the second line follows from the first simply by changing the variable
$s$ to the variable $\rho=1/s$. Thus, the maximizing $\rho$ is the negative
local slope of the function $D_Q(R)$. It follows that in the curvy part of 
$\calE_{CKM}(R,Q)$, where the slope of $D_Q(R)$ is smaller than $-1$, the
maximizing $\rho$ is larger than $1$. Thus, the maximization in the last
expression of $D_Q(R)$ can be confined to the range $[1,\infty)$, i.e.,
for $R \le R_1$
\begin{eqnarray}
\label{ECKM2}
\calE_{CKM}(R,Q)
&=&\sup_{\rho\ge
1}\left[-\rho\sum_{x\in\calX}q(x)\ln\left(\sum_{x'\in\calX}q(x')e^{-d_B(x,x')/\rho}\right)-\rho
R\right]\nonumber\\
&=&\sup_{\rho\ge
1}\left\{-\rho\sum_{x\in\calX}q(x)\ln\left(\sum_{x'\in\calX}q(x')
\left[\sum_{y\in\calY}\sqrt{p(y|x)p(y|x')}\right]^{1/\rho}\right)-\rho
R\right\}.
\end{eqnarray}
and of course, for $R\in[R_1,R_1+D_Q(R_1)]$ we use the same expression,
setting $\rho=1$. This should now be compared with Gallager's expression
\begin{equation}
\calE_G(R,Q)=
\sup_{\rho\ge 1}
\left\{-\rho\ln\left(\sum_{x,x'\in\calX}q(x)q(x')
\left[\sum_{y\in\calY}\sqrt{p(y|x)p(y|x')}\right]^{1/\rho}\right)-\rho
R\right\}.
\end{equation}
As can be seen,
the only difference between the two expressions is that in $\calE_{CKM}(R,Q)$,
the averaging over $x$ is external to the logarithmic function, whereas in
$\calE_G(R,Q)$ it is internal to the logarithmic function. Thus, Jensen's
inequality guarantees that $\calE_{CKM}(R,Q)\ge
\calE_G(R,Q)$, and since the logarithmic function is strictly concave,
the inequality is strict for every finite $\rho$ (which means $R> 0$), unless
$\sum_{x'\in \calX}q(x')e^{-d_B(x,x')/\rho}$ happens to be independent of $x$,
which is the case when either $Q$ and $P$ exhibit enough symmetry, or when
$Q$ is chosen to be the optimum distribution \cite[p.\ 193, Problem 23b, hint
(iii)]{CK81}.

Our second preliminary observation is the following. The derivation of Gallager's expurgated
exponent begins from the union bound on the pairwise error probabilities, which
in turn are all upper bounded by the Bhattacharyya bound, i.e., eq.\ (5.7.3)
in \cite{Gallager68} reads
\begin{equation}
P_{e|m}\le \sum_{m'\ne m}\sum_{\by}\sqrt{p(\by|\bx_m)p(\by|\bx_{m'})},
\end{equation}
where $\by\in\calY^n$ designates the channel output vector. 
One might suspect that a better result can probably be obtained by
considering, more generally, the Chernoff bound
\begin{equation}
P_{e|m}\le \sum_{m'\ne
m}\sum_{\by}p^s(\by|\bx_{m'})p^{1-s}(\by|\bx_m),~~~~~~0\le s\le 1,
\end{equation}
where the Chernoff parameter $s$ is subjected to optimization (in addition to
the parameter $\rho$). After carrying out the derivation similarly as in
\cite[Section 5.7]{Gallager68}, one would obtain a similar expression as in
$\calE_G(R,Q)$, except that the Bhattacharyya distance function is replaced,
more generally, by the Chernoff distance function
\begin{equation}
d_s(x,x')=-\ln\left[\sum_{y\in\calY}p^{1-s}(y|x)p^s(y|x')\right].
\end{equation}
Thus, $E_G(\rho,Q)$ would be replaced by
\begin{equation}
E_G(\rho,s,Q)=
-\rho\ln\left(\sum_{x,x'\in\calX}q(x)q(x')
\left[\sum_{y\in\calY}p^{1-s}(y|x)p^s(y|x')\right]^{1/\rho}\right),
\end{equation}
and the best choice of $s$ would be the one that maximizes $E_G(\rho,s,Q)$.
However, it is easy to see that $E_G(\rho,s,Q)$ is concave in $s$ and that
$E_G(\rho,s,Q)=E_G(\rho,1-s,Q)$ since $x$ and $x'$ play symmetric roles in the
expression of $E_G(\rho,s,Q)$. Thus, the maximizing $s$ is obviously
$s^*=1/2$, which brings us back to the Bhattacharyya distance, and confirming that
there is nothing to gain from the optimization over $s$ beyond Gallager's
expurgated bound.

This is not the case, however, when it comes to the CKM expurgated bound.
In particular, Csisz\'ar and K\"orner also begin from the union--Bhattacharyya bound (see
\cite[p.\ 186, top]{CK81}), and an extension of their derivation would
yield the same expression as (\ref{ECKM2}), but again, with the
Bhattacharyya distance $d_B(x,x')$ (or $d_{1/2}(x,x')$) 
being replaced by the more general Chernoff distance $d_s(x,x')$.
However, here $x$ and $x'$ do {\it not} have symmetric roles and hence the
bound is not necessarily optimized at $s=1/2$. Indeed, it is easy to study
a simple example of a binary non--symmetric channel
and see that the
derivative of the function
\begin{equation}
\label{ErsQ}
E(\rho,s,Q)=-\rho\sum_{x\in\calX}q(x)\ln\left(\sum_{x'\in\calX}q(x')
\left[\sum_{y\in\calY}p^{1-s}(y|x)p^s(y|x')\right]^{1/\rho}\right)
\end{equation}
with respect to $s$ does not vanish at $s=1/2$ unless $Q$ is symmetric
(see also Example 1 below, at the end of this section).

To summarize, we observe that the CKM expurgated bound is not only better, in
general, than the Gallager expurgated bound, but moreover, it provides even further room
for improvement in the optimization over $s$, in addition to the optimization
over $\rho$. Confining the framework to finite alphabets and fixed composition
codes, this gives rise to the following coding theorem.

\begin{theorem}
For an arbitrary DMC, there exist a sequence of codes $\{\calC_n\}_{n\ge 1}$ of rate $R$ and
composition $Q$,\footnote{A sequence of codes with composition $Q$ means a
sequence of fixed composition codes, where the common empirical distribution of all
codewords tends to $Q$ as $n\to\infty$.}
for which the error exponent associated with the maximum error probability
is at least as large as
\begin{equation}
\calE(R,Q)=\sup_{\rho\ge 1}\sup_{0\le s\le 1}[E(\rho,s,Q)-\rho R]
\end{equation}
where $E(\rho,s,Q)$ is defined as in eq.\ (\ref{ErsQ}).
\end{theorem}

\vspace{0.2cm}

\noindent
{\it Example 1 -- binary input, binary output channels.}
We have compared numerically the three expurgated exponents for various combinations of
$P$ and $Q$ associated with binary input, binary output channels.
As a representative example, we have computed $E_G(1,Q)$, $E(1,1/2,Q)$ and
$\max_{0\le s\le 1}E(1,s,Q)$, for the binary channel $P$ defined by
$p(0|0)=p(1|0)=0.5$, $p(0|1)=1-p(1|1)=10^{-10}$, along with the input assignment $Q$
given by
$q(1)=1-q(0)=0.1$. The results are $E_G(1,Q)=0.0542$, $E(1,1/2,Q)=0.0574$, and
$\max_{0\le s\le 1}E(1,s,Q)=0.0596$, which is achieved at $s^*\approx 0.76$.
This means that in the range of high rates, we have
\begin{eqnarray}
\calE_G(R,Q)&=&0.0542-R\\
\calE_{CKM}(R,Q)&=&0.0574-R\\
\calE(R,Q)&=&0.0596-R.
\end{eqnarray}
Thus, numerical evidence indeed supports the fact that there are gaps between the
three expurgated exponents, at least for some combinations of channels and
input assignments.

\section{New Derivation of the Expurgated Exponent}

Equipped with the background of Section 2 and the observations offered in
Section 3, we next proceed to the derivation of the new version of the
expurgated bound (i.e., prove Theorem 1), but in a manner that does not
rely on the packing lemma and hence is not sensitive
to the assumptions of fixed composition codes and finite alphabets. 
We will assume finite alphabets only for the simplicity of the
exposition and for the sake
convenience, but it should be understood that our analysis has a natural
extension to continuous alphabets (along with channel input constraints), and
we will outline this extension in Section 5.

Following the discussion in Section 3, we begin with the following upper bound
on the conditional probability of error
\begin{equation}
P_{e|m}\le \sum_{m'\ne
m}\sum_{\by}p^s(\by|\bx_{m'})p^{1-s}(\by|\bx_m),~~~~~~0\le s\le 1.
\end{equation}
Now, following the same rationale as in \cite[Section 5.7]{Gallager68} 
and \cite[Section 3.3]{VO79}, we argue the following:
There exists a codebook $\calC_n=\{\bx_1,\ldots,\bx_M\}$ of
$M=e^{nR}$ codewords such that for every $\rho > 0$ and all $1\le m \le M$
\begin{equation}
P_{e|m} \le \left[2\overline{P_{e|m}^{1/\rho}}\right]^\rho\le
2^\rho\left[\bE\left(\sum_{m'\ne
m}\sum_{\by}p^s(\by|\bX_{m'})p^{1-s}(\by|\bX_{m})\right)^{1/\rho}\right]^\rho \dfn
2^{\rho}A_n(R,\rho),
\end{equation}
where the expectation operator is taken w.r.t.\ the randomness of the
codewords $\{\bX_m\}$, which are selected independently at random
according to the uniform distribution over the type class $T_Q$, that is, the
set of all sequences whose empirical distribution is (as close as possible to)
$Q$. 

For the purpose of further bounding $A_n(R,\rho)$,
the next step in both \cite{Gallager68} and \cite{VO79} is to use the
inequality $[\sum_{m'}a_{m'}]^{1/\rho}\le\sum_{m'}a_{m'}^{1/\rho}$, which
holds for every
$\rho\ge 1$, and then to apply the expectation operator 
on each term of the corresponding sum
separately. This is a step which simplifies the derivation to a large extent,
but at the possible price of losing exponential tightness of the resulting bound.
Instead, in our derivation, we will use another approach, which yields
an exponentially tight bound.
Defining
\begin{equation}
d_s(x,x')=-\ln\left[\sum_yp^{1-s}(y|x)p^s(y|x')\right]
\end{equation}
we have, due to the memorylessness of the channel,
\begin{equation}
\sum_{\by}p^{1-s}(\by|\bx_m)p^s(\by|\bx_{m'})=e^{-\sum_{i=1}^nd_s(x_{m,i},x_{m',i})}\dfn
e^{-d_s(\bx_m,\bx_{m'})},
\end{equation}
where $x_{m,i}$ is the $i$--th component of the codeword $\bx_m$.
Let $N_m(\hat{Q}_{XX'})$ be the number of codewords $\{\bx_{m'}\}$ that,
together with $\bx_m$, fall in the joint
type class corresponding to the joint empirical distribution $\hat{Q}_{XX'}$,
whose both marginals must agree with $Q$ (as they are both empirical
distributions of codewords).
Then, we have
\begin{eqnarray}
\label{longchain}
A_n(R,\rho)&=&\left[\bE\left(\sum_{\hat{Q}_{XX'}}N_m(\hat{Q}_{XX'})
\exp\{-n\bE d_s(X,X')\}\right)^{1/\rho}
\right]^\rho\nonumber\\
&\exe&\left[\bE\left(\max_{\hat{Q}_{XX'}}N_m(\hat{Q}_{XX'})
\exp\{-n\bE d_s(X,X')\}\right)^{1/\rho}
\right]^\rho\nonumber\\
&=&\left[\bE \max_{\hat{Q}_{XX'}}[N_m(\hat{Q}_{XX'})]^{1/\rho}
\exp\{-n\bE d_s(X,X')/\rho\}
\right]^\rho\nonumber\\
&\exe&\left[\bE \sum_{\hat{Q}_{XX'}}[N_m(\hat{Q}_{XX'})]^{1/\rho}
\exp\{-n\bE d_s(X,X')/\rho\}
\right]^\rho\nonumber\\
&=&\left[\sum_{\hat{Q}_{XX'}}\bE\left\{[N_m(\hat{Q}_{XX'})]^{1/\rho}\right\}\cdot
\exp\{-n\bE d_s(X,X')/\rho\}
\right]^\rho\nonumber\\
&\exe&\left[\max_{\hat{Q}_{XX'}}\bE\left\{[N_m(\hat{Q}_{XX'})]^{1/\rho}\right\}\cdot
\exp\{-n\bE d_s(X,X')/\rho\}
\right]^\rho\nonumber\\
&\exe&\max_{\hat{Q}_{XX'}}\left(\bE\left\{[N_m(\hat{Q}_{XX'})]^{1/\rho}\right\}\right)^\rho\cdot
\exp\{-n\bE d_s(X,X')\},
\end{eqnarray}
where the notation $\exe$ designates equivalence in the exponential scale (i.e.,
$a_n\exe b_n$ means that $\frac{1}{n}\ln\frac{a_n}{b_n}\to 0$ as
$n\to\infty$), and where the expectation at the exponent is w.r.t.\
$\hat{Q}_{XX'}$.
Now, similarly as in \cite[p.\ 4444, eq.\ (34)]{p119}, we have
\begin{equation}
\label{moment}
\bE\left\{[N_m(\hat{Q}_{XX'})]^{1/\rho}\right\}\exe\left\{\begin{array}{ll}
\exp\{n[R-I(X;X')]\} & R < I(X;X')\\
\exp\{n[R-I(X;X')]/\rho\} & R \ge I(X;X')\end{array}\right.
\end{equation}
where $I(X;X')$ is the mutual information between $X$ and $X'$ associated with
$\hat{Q}_{XX'}$. This result follows from the fact that given $\bX_m=\bx_m$,
$N_m(\hat{Q}_{XX'})$ is the sum of $e^{nR}-1$ binary independent
random--variables, 
\begin{equation}
U_{m'}=1\{(\bx_m,\bX_{m'})~\mbox{have empirical joint
distribution}~\hat{Q}_{XX'}\},~~~m'\ne m,
\end{equation}
whose expectations are all of the exponential order of $e^{-nI(X;X')}$.
Upon taking into account all the possible empirical distributions
$\{\hat{Q}_{XX'}\}$, we readily obtain
\begin{equation}
A_n(R,\rho)\exe e^{-n\min\{E_1(R),E_2(R,\rho)\}}, 
\end{equation}
where
\begin{equation}
E_1(R,\rho)=\min_{\hat{Q}_{X'|X}:~I(X;X')\ge R}[Ed_s(X,X')+\rho
I(X;X')]-\rho R
\end{equation}
and
\begin{equation}
E_2(R)=\min_{Q_{X'|X}:~I(X;X')\le R}[Ed_s(X,X')+
I(X;X')]-R=\sup_{\rho\ge 1}[E(\rho,s,Q)-\rho R],
\end{equation}
where the second equality is obtained similarly 
as in the derivation of eq.\ (\ref{ECKM2}), but with
the Bhattacharyya distortion measure being replaced by $d_s(\cdot,\cdot)$.
It remains to show that $E_1(R,\rho)$, for the optimum choice of $\rho$, is
never smaller than $\calE_{CKM}(R,Q)$.
For a given $s$, let $R_Q(D)$ be the rate--distortion function of $X$ w.r.t.\ the distortion
measure $\{d_s(x,x')\}$ subject to the constraint that $Q_{X'}=Q$.
Let $D_\rho$ be the distortion level at which $R_Q'(D)=-1/\rho$, where
$R_Q'(\cdot)$ is the derivative of $R_Q(\cdot)$. Also, $D_Q(R)$ will denote
the corresponding distortion--rate function, which is the inverse of $R_Q(D)$.
Then $E_1(R,\rho)$ admits the following expressions:
\begin{equation}
E_1(R,\rho)=\left\{\begin{array}{ll}
D_\rho+\rho[R_Q(D_\rho)-R] & R\le R_Q(D_\rho)\\
D_Q(R) & R\ge R_Q(D_\rho)\end{array}\right.
\end{equation}
As the straight line $D_\rho+\rho[R_Q(D_\rho)-R]$ is tangential to (and below)
the convex function $D_Q(R)$, the best choice of $\rho$ is to take the limit $\rho\to\infty$.
But $E_1(R,\infty)=D_Q(R)$ for all $R$ (as $R_Q(D_{\infty})=0$), which is in turn
at least as large as $E_2(R)=\sup_{\rho\ge 1}[E(\rho,s,Q)-\rho R]$ for all
$R$, and strictly so in the linear part of the latter function.

Thus, for a given $s$, there exists a sequence of codes for which the exponent of the 
maximum probability of error is dominated by $\sup_{\rho\ge
1}[E(\rho,s,Q)-\rho R]$. Upon maximization over $s$, this yields
$\calE(R,Q)$, as asserted in Theorem 1.

\section{Beyond Finite Alphabets and Fixed Composition Codes}

In Section 4, we have assumed finite alphabets and fixed composition codes,
mainly for the simplicity of the exposition and for the purpose of comparison
with the CKM expurgated exponent. However, as we have mentioned already, the
analysis in Section 4 is not really sensitive to these assumptions.

The heart of the analysis in Section 4 is around equations (\ref{longchain})
and (\ref{moment}), and therefore, the
main issue in the desired extension is to adapt this
part of the analysis to continuous alphabets. 
Consider now the case where $\calX=\calY=\reals$ and then
$q(x)$ and $p(y|x)$ are probability density functions.
Let $\delta$ be an arbitrarily small positive real. Then,
\begin{equation}
\sum_{m'\ne m} e^{-d_s(\bx_m,\bx_{m'})}\le\sum_{k=0}^\infty
e^{-nk\delta}N_m(k),
\end{equation}
where
\begin{equation}
N_m(k)=\sum_{m'\ne m} 1\left\{nk\delta\le d_s(\bx_m,\bx_{m'})<
n(k+1)\delta)\right\},~~~~k=0,1,2,\ldots
\end{equation}
Let us assume now that the ensemble of codes is defined such that
$d(\bx_m,\bx_{m'})$ cannot
exceed $nD_{\max}$, where $D_{max} < \infty$ is a constant that does not
depend on $n$, which is normally the case when the codewords must comply with
input constraints. Then
using a similar technique as in eq.\ (\ref{longchain}), 
we now obtain
\begin{equation}
A_n(R,\rho)\lexe \sup_{k\ge 0}
\left(\bE\{[N_m(k)]^{1/\rho}\}\right)^\rho\cdot e^{-nk\delta},
\end{equation}
where the notation $\lexe$ denotes inequality in the exponential scale (more
formally, $a_n\lexe b_n$ means $\limsup_{n\to\infty}\frac{1}{n}\ln
\frac{a_n}{b_n}\le 0$). The key issue is now to assess the exponential rate of
the expectation of the binary random variable, 
\begin{equation}
U_{m'}=1\left\{nk\delta\le d_s(\bx_m,\bX_{m'})<
n(k+1)\delta\right\},
\end{equation}
for a given $\bx_m$, namely, to find the exponent of 
$\mbox{Pr}\{nk\delta\le d_s(\bx_m,\bX_{m'})<
n(k+1)\delta\}$. This can be done using standard large deviations techniques,
like the Chernoff bound. Let $R(k\delta)$ denote the large deviations rate
function of this probability (which depends, of course, on $\bx_m$, but
it would be convenient to define the ensemble such that this rate function will be
the same for all $m$). Then, as
in eq.\ (\ref{moment}), we then have
\begin{equation}
\bE\{[N_m(k)]^{1/\rho}\}\exe\left\{\begin{array}{ll}
\exp\{n[R-R(k\delta)]\} & R \le R(k\delta)\\
\exp\{n[R-R(k\delta)]/\rho\} & R > R(k\delta)\end{array}\right.
\end{equation}
Now, similarly as in Section 4, $A_n(R,\rho)$ is dominated by
$\min\{E_1(R,\rho,\delta),E_2(R,\delta)\}$, where
\begin{equation}
E_1(R,\rho,\delta)=\inf_{k:~R(k\delta)\ge R}[k\delta+\rho R(k\delta)]-\rho R,
\end{equation}
and
\begin{equation}
E_2(R,\delta)=\inf_{k:~R(k\delta)\le R}[k\delta+R(k\delta)]-R,
\end{equation}
Upon taking the limit $\delta\to 0$, these become
\begin{equation}
E_1(R,\rho)=\inf_{D:~R(D)\ge R}[D+\rho R(D)]-\rho R,
\end{equation}
and
\begin{equation}
E_2(R)=\inf_{D:~R(D)\le R}[D+R(D)]-R.
\end{equation}
The remaining details depend, of course, on the form of the large
deviations rate function
$R(D)$, which in turn depends strongly on the input assignment and the
channel.

\vspace{0.2cm}

\noindent
{\it Example 2 -- the Gaussian channel.}
Consider the memoryless additive Gaussian channel $Y=X+Z$, where $Z$
is a zero--mean Gaussian random variable with variance $\sigma^2$, independent of $X$.
Let $q(\bx)$ be the uniform distribution over the surface of the
$n$--dimensional sphere with radius $\sqrt{nS}$. In this case, 
the Chernoff distance is maximized at $s^*=1/2$, where it agrees with
the Bhattacharyya distance $d_B(x,x')=(x-x')^2/8\sigma^2$.
It is not
difficult to show (e.g., using the methods of \cite{Merhav93}) that
\begin{equation}
R(D)=\frac{1}{2}\ln\left[\frac{S}{8\sigma^2D(1-2\sigma^2D/S)}\right],
\end{equation}
which has the interpretation of the rate--distortion function of the Gaussian
source with variance $S$ w.r.t.\ Bhattacharyya distortion measure with the additional
constraint that reproduction variable $X'$ is also Gaussian, zero--mean and
with variance $S$. The corresponding distortion--rate function (which is the inverse of
$R(D)$) is given by
\begin{equation}
D(R)=\frac{S(1-\sqrt{1-e^{-2R}})}{4\sigma^2},
\end{equation}
which is also the curvy part of the corresponding expurgated exponent. The
linear part is again the tangential straight line with slope $-1$.

\section{The Statistical--Mechanical Perspective}

Let us take another look at the central expression that was handled in
Sections 4 and 5, namely, on the summation
\begin{equation}
\label{REM}
Z=\sum_{m'\ne m} e^{-d_s(\bx_m,\bx_{m'})},
\end{equation}
From the viewpoint of statistical physics, this can be interpreted as the
partition function of a physical system, where for a fixed $\bx_m$, the
various configurations
(microstates) are $\{\bx_{m'}\}_{m'\ne m}$ and
the Hamiltonian (energy function)
is given by (or proportional\footnote{To enhance the analogy with physics, it is
instructive to consider a parametric family of channels, $p_\beta(y|x)\propto
[p(y|x)]^\beta$, where $\beta$ is a parameter that controls the `quality' of
the channel (e.g., the SNR in the case of the Gaussian channel), 
whose physical meaning is inverse temperature. In this case,
$d_s(\bx_m,\bx_{m'})$ of the channel pertaining to $\beta=1$ 
would be multiplied by $\beta$, similarly as in ordinary
partition functions.}
to) $d_s(\bx_m,\bx_{m'})$. If the correct codeword $\bx_m$
is given and the remaining codewords are considered independent and random,
thus denoted $\{\bX_{m'}\}$, then the various ``configurational energies''
$\{d_s(\bx_m,\bX_{m'})\}$ are also independent random variables. As explained
in \cite[Chapters 5, 6]{MM09} (see also \cite[Chapters 6, 7]{p138} and
references therein), this setting is analogous to the random energy model (REM) in
the literature of statistical physics of magnetic materials. The REM
was invented by Derrida \cite{Derrida80a}, \cite{Derrida80b},
\cite{Derrida81} as a model of extremely disordered spin glasses.
This model is not realistic, but it is exactly solvable and it exhibits a 
phase transition: Below a certain critical temperature, the partition function becomes
dominated by a sub--exponential number of configurations,
which means that
the system freezes in the sense that its
entropy vanishes in the thermodynamic limit. This combination of freezing and
quenched disorder resembles the behavior of a glass, and so, this low temperature
phase of zero entropy is called the {\it glassy phase}.\footnote{In physics,
it typically occurs as a result of a process of rapid cooling.}
Above the critical temperature, the partition function is dominated by an
exponential number of configurations, and so, its entropy is positive. This
high temperature phase is called the {\it paramagnetic phase.}

In the derivations of Sections 4 and 5, the curvy part of the graph of
$\calE(R,Q)$ corresponds to the glassy phase of the REM associated with
(\ref{REM}), because the dominant contribution to $A_n(R,\rho)$ is due to
a subexponential number ($N_m(\hat{Q}_{XX'})$ or $N_m(k)$) of codewords
whose distance from $\bx_m$ is about $nD_Q(R)$. The straight--line part of
$\calE(R,Q)$, on the other hand, corresponds to the paramagnetic phase, where
about $e^{n[R-R_1]}$ incorrect codewords at distance $nD_Q(R_1)$ dictate the
behavior. Thus, the passage between the curvy part and the straight--line
part, at $R=R_1$ is interpreted as a glassy phase transition.

In the Gallager expurgated bound, there is also a passage from a
curvy part at low rates to a straight--line part at high rates. However, in
Gallager's derivation, the passage happens due to a more 
technical reason. Since
Gallager's analysis is based on the inequality
$[\sum_{m'}a_{m'}]^{1/\rho}\le\sum_{m'}a_{m'}^{1/\rho}$, which holds only for
$\rho\ge 1$, the 
maximization over $\rho$ is a--priori limited to the range
$\rho\ge 1$. The linear part of the curve is then generated due to the
fact that for higher rates,
the unconstrained achiever of $E_{ex}(R)$ is $\rho^*< 1$, and so,
the constrained one remains $\rho^*=1$, independently of $R$ in this range.



\end{document}